\documentclass{raa}            

\usepackage{graphicx,times}             

\begin{document}

  \title{A solution of the puzzling symbiotic X-ray system 4U 1700+24}

   \volnopage{Vol.0 (200x) No.0, 000--000}      
   \setcounter{page}{1}          

     \author{Ren-Xin Xu
   }

   \institute{School of Physics and State Key Laboratory of Nuclear Physics and Technology, and\\
Kavli Institute for Astronomy and Astrophysics, Peking University, Beijing 100871, P. R. China
           }

   \date{Received~~2012 month day; accepted~~2012~~month day}

\abstract{%
A circumstellar corona is proposed for a strange quark-cluster star during an accretion phase, that could be essential to understand the observations of the puzzling symbiotic X-ray system, 4U 1700+24.
The state of cold matter at supranuclear density is still an important matter of debate, and one of the certain consequences of strange star as the nature of pulsars is the self-bound on surface which makes extremely low-mass compact objects unavoidable.
In principle, both the redshifted O VIII Ly-$\alpha$ emission line and the change of the black-body radiation area could be understood naturally if 4U 1700+24 is a low-mass quark-cluster star in case of wind-accretion.
}

   \authorrunning{Xu}            
   \titlerunning{A solution of 4U 1700+24}  

   \maketitle

%
%
\section{Introduction}           

It has been more than 80 years since the first idea of stars formed of matter at nuclear (even supranuclear) density (see summaries and new achievements edited by van Leeuwen (2013) for IAU Symposium 291), but (un)fortunately, scientific challenges and opportunities still exit, especially after the discoveries of massive pulsars (Demorest et al. 2010, Antoniadis et al. 2013).
Great efforts, though difficult due to unknown microphysics, are tried to model the inner structure of pulsar-like compact stars.
Traditionally, quarks are believed to be confined in hadrons of neutron stars, while quark matter would exist in a core of hybrid/mixed star whose central density could be high enough to make quarks de-confined, and a radical but logical argument is that an entire compact star could be composed of quark matter, called a quark star.
In addition to all of the above, among different models of pulsar structure, a quark-cluster star is a condensed object of quark clusters via residual color interaction, which distinguishes from conventional both neutron and quark stars (Xu 2010, 2013).

Hadron star and hybrid/mixed star are bound by gravity, which are covered by crusts with nuclei and electrons, whereas quark star and quark-cluster star are strongly self-bound on surface.
It is then unavoidable that the mass of strange quark/quark-cluster star could be extremely low, even as low as planet-mass (so called strange planet; Xu \& Wu 2003, Horvath 2012), although massive quark-cluster stars as high as $\sim 3M_\odot$ are still possible because of a very stiff equation of state of quark-cluster matter (Lai \& Xu 2009, Lai, Gao \& Xu 2013).
Certainly, besides massive pulsars, it is also necessary to find evidence for low-mass compact objects in order to test the quark-cluster model. We would address in this paper that the discovery of the redshift (Tiengo et al. 2005, confirmed recently by Nucita et al. (2014) with publicly available X-ray data) from 4U 1700+24 could be one of such attempts.

In fact, it has been a long history to search for atomic transition from spectral features of either none-accreting (e.g., Burwitz et al. 2001) or accreting (e.g., Cottam et al. 2002) compact stars, but we have not yet seen success with certainty previously (e.g., Kong et al. 2007).
The detected redshift may imply an emission cloud at a distance of $160-1000$ km from the central object if one assumes a neutron star mass of $\sim 1.4 M_\odot$ for 4U 1700+24 (Nucita et al. 2014).
However, our solution of the system in the regime of quark-cluster star is that the central object has an extremely low mass of only $\sim 10^{-2} M_\odot$.

We will show that a corona-like plasma may exist above a quark-cluster star, and atomic transition lines could form in the corona if, in a high accretion phase, its density is larger than a critical value.
In the limit of $z\ll 1$, the gravitational redshift, $z$, of a compact star reads (e.g., Xu 2003),
\begin{equation}
z\simeq {GM\over c^2R},
\end{equation}
for a stellar mass of $M$ and a radius of $R$. For low mass quark/quark-cluster stars, $M\ll M_\odot$, their densities are nearly uniform, so we have
\begin{equation}
R=\sqrt{{3c^2\over 4\pi G\rho_\star}z},~~M={4\pi\over 3}R^3\rho_\star,%
\label{MR}
\end{equation}
where $\rho_\star$ is the uniform density of quark-cluster star at pressure-free state. We are applying $\rho_\star=2\rho_0$ in following calculations (e.g., Lai \& Xu 2009), with $\rho_0$ the nuclear saturation density.
For $z=0.009$, we have $R=2.3$ km and $M=0.014M_\odot$, which hints that the compact object could be an extremely low-mass quark-cluster star, being gravity-free (see Fig. 1 of Xu 2010).

\section{The model}

\subsection{Polar corona of an accreting strange quark-cluster star}

For a pure strange quark star, on one hand, if the kinematic energy of ions above quark surface is zero (ie., without inclusion of barrier penetration), the Coulomb force formulated in Eq.(\ref{barrier}) could support a crust with mass as much as $\sim 10^{-5}M_\odot$ (Alcock et al. 1986; unfortunately, the authors concluded that the transmission probability of $\sim 10^{-104}$ is negligible if $A=118$ and $Z=36$, but the probability is $\sim 10^{-18}$ for $^{16}$O with the same approximations presented).
However, on the other hand, with the inclusion of barrier penetration, ions even at thermal energy equilibrium may significantly penetrate (see a discussion in \S~2.2) and be consequently deconfined into quark matter (Note: the kinematic energy of a free-fall ion is usually higher than $\sim 10$ MeV).

Note that things are very different for a strange quark-cluster star.
We will demonstrate here that, when accretion rate is much lower than the Eddington limit ($\rho$-low case, see below), a quark-cluster star would encompass corona-like ions above polar cap.
If the accretion rate is high ($\rho$-high case), then the corona ions may diffuse across magnetic field lines, spreading over almost all the stellar surface.
A key point in this scenario is: although ions may penetrate significantly the barrier, most of them would bounce back into corona again.
In addition, we would expect a two temperature system of the corona, where the temperature of ions, $T_{\rm i}$, should generally be much higher than that of electron, $T_{\rm e}$, because an electron in strong magnetic field could lost its energy very effectively (while not for ions).
The thermal X-rays observed are related to electron temperature, $T_{\rm e}$.

Because of strong electron screening effect, the Coulomb barrier for an ion with charge number, $Z$, is (Xu et al. 1999)
\begin{equation}
V_{\rm Coul} = {3ZV_{\rm q}\over \sqrt{6\alpha/\pi}V_{\rm q}\hat{z}/(\hbar c)+4}\propto {1\over \hat{z}}, ~~~({\rm when} \hat{z}\gg 10^2 {\rm fm})%
\label{barrier}
\end{equation}
where $\hat{z}$ is the height above the stellar surface, $\alpha=e^2/\hbar c$ is the fine-structure constant, and $V_{\rm q}^3/(3\pi^2\hbar^3c^3)$ is the charge density of quark-clusters. It is evident, fortunately, that the Coulomb potential looks like that of point charge rather than of 1-dimensional electric capacity, so that we could apply some textbook conclusions (e.g., Gamow energy below) of nuclear reaction for some approximations.
The value of $V_{\rm q}$ dependents on light-flavor symmetry breaking of strange matter, that is difficult to calculate exactly, but would be order of 10 MeV.

We could approach the physics of polar corona at two limits: high ($\rho$-high) and low ($\rho$-low) density cases, which would be separated by a critical density of corona ions,
\begin{equation}
\rho_{\rm c} \simeq {Am_{\rm u}\over r_{\rm L}^3}\sim 5.5\times 10^{-8}\beta^{-3}A^{-2}Z^3B_{12}^3 ~~~{\rm g/cm^3},
\end{equation}
where the Larmor radius of an ion ($A,Z$) in a constant magnetic field $B=B_{12}\times 10^{12}$ reads,
\begin{equation}
r_{\rm L}=\beta{Am_{\rm u}c^2\over ZeB}\sim 3.1\times 10^{-6}\beta {A\over Z}B_{12}^{-1}~{\rm cm}.
\end{equation}%
\label{rL}
For $v_{\rm f}=0.1c$ (i.e., $\beta\equiv v/c=0.1$) in Eq.(\ref{vf}), $A/Z=2$, $Z=10$, $B_{12}=1$, one has $\rho_{\rm c}\sim 10^{-4}$ g/cm$^4$ (i.e., ion number density of $\sim 10^{18}$ cm$^{-3}$, approximately the density of air).

In the $\rho$-low case of $\rho < \sim \rho_{\rm c}$, the ion plasma could be collision free, with a temperature $kT_{\rm i}\sim E_{\rm g}\sim 10$ MeV (see Eq.(\ref{Eg}) in \S~2.2). All the falling ions could penetrate the Coulomb barrier, but may bounce back into outside almost along the magnetic field lines, forming a corona with energetic ions, because of very low rate of weak interaction to change up and down to strange quarks.
Bondi accretion from interstellar medium may power the thermal X-ray emission of a dead pulsar, like the ``Magnificent Seven'' (e.g., Mereghetti 2011) with X-ray luminosity $L_{\rm x}\sim 10^{30-31}$ erg/s.
The density of ion in corona could be estimated as
\begin{equation}
\rho_{\rm Bondi} \sim {L_{\rm x}\over \eta c^2 v_{\rm f} r_{\rm p}^2}({r_{\rm p}\over R'})^2\sim 10^{-10}\eta^{-1}~~{\rm g/cm^3},%
\label{rho-Bondi}
\end{equation}
for $v_{\rm f}\simeq 0.1c$ in Eq.(\ref{vf}) and polar ions diffuse over a star surface with radius $R'\sim 10^5$ cm. The density could be too low to result atomic feature lines in thermal X-ray spectra if the probability $\eta > 10^{-10}$ (see \S~2.2 for a discussion of $\eta$-value). Also the polar hot spot could not be clear because of low temperature gradient and high thermal conductivity inside compact star.

In the $\rho$-high case of $\rho\gg \rho_{\rm c}$, ion collision dominates, and a falling ion may lost quickly both kinematic energies perpendicular and parallel to the field lines. The ion plasma temperature $kT_{\rm i}>\sim$ keV would be much lower than $E_{\rm g}$. This is the case where significant accretion occurs, via either Roche lobe overflow or wind accretion.
As for the X-ray spectrum, atomic features could form in this case.
In addition, a massive quark-cluster star may manifest as an X-ray burster even a sperburster when its strain energy develops inside the solid star and reaches a critical value, i.e., a quake occurs, because significant gravitational and elastic energies release.
Certainly, more detailed researches are necessary for this scenario, without the cool-crust challenge in conventional X-ray burst model of neutron star (Altamirano et al. 2012).

As for the composition of ions in the corona, nuclei with high nuclear binding energy are expected there, otherwise they would be destroyed due to frequent collisions with quark-clusters. Most of the nuclei could then be $\alpha$-nuclei (e.g., $^{16}$O, $^{56}$Fe, etc.) and proton/neutrons in the corona.

\subsection{An order-of-magnitude understanding of the polar corona}

The kinematic energy of an ion (with atomic number $A$) accreted onto the surface would be
\begin{equation}
E_{\rm g} \sim {GAMm_{\rm u}\over R}\simeq {4\pi\over 3}Gm_{\rm u}\rho A R^2 \simeq 8.4A~~ {\rm MeV},%
\label{Eg}
\end{equation}
with a velocity of
\begin{equation}
v_{\rm f}=\sqrt{2GM\over R}=\sqrt{{8\pi\over 3} G\rho}~R\simeq 0.13 c,%
\label{vf}
\end{equation}
where $m_{\rm u}$ is the atomic mass unit and stellar radius $R$ is suggested to be a few kilometers.

Let's first consider the polar cap of an accreting quark-cluster star. In an environment of radiation with electron temperature $T_{\rm e}\sim 1$ keV, atoms of $Z<10$ would be totally ionized, becoming a gas of free nuclei and electrons.
Because electrons feel a radiation pressure outward, and a radiation-induced electric field will form.
The radiation force should be balanced by an electric one, $E_{\rm r}$,
\begin{equation}
a\sigma_{\rm T}T_{\rm e}^4\simeq eE_{\rm r},
\end{equation}
with $a$ the radiation density constant and $\sigma_{\rm T}$ the Thomson cross section. One then has $E_{\rm r} \simeq a\sigma_{\rm T}T^4/e \simeq 3\times 10^{-3} T_6^4$ V/cm, where $T_6=T_{\rm e}/10^6$K, which is really negligible compared with the electric field $E_{\rm s}\sim 10^{17}$ V/cm on strange matter surface (e.g., Xu \& Qiao 1999), both directed outward.

Although $E_{\rm r}\ll E_{\rm s}$, it doesn't means that the radiation-induced electric field, $E_{\rm r}$, could not halt significantly an accreted ion falling onto the polar cap.
The length-scale of decelerating ion by $E_{\rm s}$ is order of $10^3$fm, while that by $E_{\rm r}$ would be order of polar cap radius, $r_{\rm p}$,
\begin{equation}
r_{\rm p}=R\arcsin\sqrt{2\pi R\over cP}\simeq 4.6\times 10^3P_{-3}^{-1/2}R_6^{3/2}{\rm m},%
\label{rp}
\end{equation}
where $P=P_{-3}\times 10^{-3}$s is the spin period, $R_6=R/10^6$cm. Therefore the electrostatic potential barrier induced by $E_{\rm s}$ could be $V_{\rm Coul}\sim 10Z$ MeV, but that by  $E_{\rm r}$ could be $V_{\rm r}\sim ZE_{\rm r} r_{\rm p}\sim 14.5ZP^{-1/2}R_6^{3/2}$eV.
We can also have $V_{\rm r}\ll V_{\rm Coul}$ even if $P=1$ms and $R_6=1$. We thus neglect the radiation-induced electric field in following calculations.

The cyclotron radiation power of an ion would be (e.g., Rybicki \& Lightman 2004),
\begin{equation}
P_{\rm c}={2\over 3}{Z^4e^4\over A^2m_{\rm u}^2c^3}\beta^2B^2\simeq 1.1\times 10^{-15}\beta^2B^2{Z^4\over A^2}({m_{\rm e}\over m_{\rm u}})^2~{\rm erg/s},
\end{equation}
for an isotropic distribution of ion velocities, $v=\beta c$. The lifetime of ion cyclotron radiation is then
\begin{equation}
\tau \simeq {Am_{\rm u}v^2 \over 2P_{\rm c}}={m_{\rm u}^3c^2\over 2.2\times 10^{-15}m_{\rm e}^2} {A^3\over Z^4B^2}\simeq 2.3\times 10^{-6}B_{12}^{-2}{A^3\over Z^4}~{\rm s}.
\end{equation}
One can conclude that corona ions may move along the magnetic field lines due to cyclotron radiation.

Let's turn to the reaction between ion and quark-cluster star. In comparison with thermal nuclear reaction (e.g., Clayton 1983), there are differences (the geometrical factor is $\sim E^{-1}$ for nuclear interactions, while it could be approximately one in our case) and similarities (occasionally weak interaction may occur in both cases to change the quark flavors).

The kinematic energy of ion (with typical temperature $T_{\rm i}$) in polar corona with height $h\ll R$ could be approximated as
\begin{equation}
 kT_{\rm i}\sim E_{\rm k} \simeq {4\pi \over 3} GAm_{\rm u}\rho R h = 1.6A{R\over {\rm km}} {h\over {\rm m}} ~{\rm keV},
\end{equation}
which could be much lower than the Gamow energy (e.g., Perkins 2009)
\begin{equation}
E_{\rm G} = ({2m\over \hbar^2})({Z_1Z_2e^2\over 4})^2 \sim 10~{\rm MeV}.
\end{equation}
The average penetration probability of one ion (note: only $\sim 10^{-5}$ part of ions are in the Gamow window, being energetic enough to possibly penetrate) is then $\sim 10^{-5}\times \exp[-\sqrt{E_{\rm G}/E_{\rm k}}]\sim 10^{-18}$.
%
%
It is then expected that, if without replenishment by accretion, all the ions would penetrate the Coulomb barrier during a timescale of $\sim 10^4$ seconds in case that the bottom ion number density of polar corona is about $10^{26}$ cm$^{-3}$ (mass density $10^2$ g/cm$^3$) due to frequent collisions (see Eq.(\ref{tauc})).

However, a successful barrier penetration doesn't mean a fusion of an atomic nucleus and the condensed matter of quark clusters because of weak interaction: either an up or a down quark has to be converted to be a strange quark at a relatively long timescale (typically at $\tau_{\rm weak}\sim 10^{-7}$ s, but sometimes as long as  $\sim 10^2$ s).
Without flavor change, nuclear fusion is the result of quick strong force, while for flavor-changed weak interaction, the cross section is suppressed significantly.
For instance, in the environment of stellar nucleosynthesis, the $S(E)$ factor is about $10^0$ keV$\cdot$barn for the reaction of C$^{12}(p,\gamma)$N$^{13}$ (Fig. 4-5 in Clayton 1983), but the $S(E)$ factor is only $\sim 10^{-22}$ keV$\cdot$barn for the flavor-changed reaction of $p+p\rightarrow d+e^++\nu_{\rm e}$ (Section 5-1 in Clayton 1983).
This hints that only negligible (at a relative probability of order 1 in $10^{20}$) number of penetrated nuclei would participate in weak interaction.
Note that the fusion of nucleus and strange matter is also via weak interaction, to create strangeness. So, in summary, most of the penetrated ions are scattered elastically via the strong interaction, forming a polar corona with height $h$. It is then reasonable to assume that the structure of corona evolves secularly at a longer time scale, and we may think that the number of weak-interaction ions equals to that of accreted ones.

It is still uncertain to calculate the relative probability, $\eta$, at which penetrated ions would successfully be flavor-changed to strange matter.
Nevertheless, one may estimate a value of $\eta\sim ~{\rm fm}/(v\tau_{\rm weak})$, from $10^{-25}$ to $10^{-13}$ if $\tau_{\rm weak}\in (10^{-7}, 10^3)$ s and $v\in(10^{-3}, 10^{-1})c$.
By an analogy of spontaneous emission of atom via electromagnetic interaction, with the inclusion of energy-dependence ($E^3$, e.g., see van Driel et al. (2005) for a test of the $E^3$-dependence) of the rate, the probability of weak interection, $\eta$, would be enhanced by a factor of $10^{3\sim 5}$ if about $(30\sim 100)$ MeV energy would be released per baryon during the phase conversion.
Note that, as quark-cluster mass is comparable to that of falling ion, part of penetrated ions would lost most of their kinematic energy during bombardment so that they could not be energetic enough to penetrate back to corona, and finally be converted to strange matter.
It is thus possible that $\eta$-value could be as high as $>10^{-5}$. This speculation cannot be ruled out now, but more elaborate work is needed to know the facts.

The charged ions can also diffuse across the magnetic field lines due to the collisions between the ions that gyrate around the lines, since a collision will alter an ion's direction and make it to gyrate around another field line.
The moving changes when the Coulomb energy approaches $kT_{\rm i}$, and the collision times scale is
\begin{equation}
\tau_{\rm c} \simeq {A^{1/2}m_{\rm u}^{1/2}(kT_{\rm i})^{3/2}\over \alpha^2\hbar^2c^2Z_1^2Z_2^2n}\sim 10^{-14} A^{1/2}Z_1^{-2}Z_2^{-2}(kT_{\rm i}/{\rm keV})^{3/2} {10^{26}{\rm cm}^{-3}\over n}~~{\rm s},%
\label{tauc}
\end{equation}
where $n$ is ion number density.
The ion diffusion across magnetic field lines could also be considered as a random work (e.g., Goldson \& Rutherford 1995). The time scale for polar corona ions to diffuse over all the stellar surface would be
\begin{equation}
\tau_{\rm diffu} \simeq ({R\over r_{\rm L}})^2\tau_{\rm c}\sim 10^3\beta A^{3/2}Z^{-5}B_{12}^{-1}(kT_{\rm i}/{\rm keV})^{3/2}R_6^2 {10^{26}{\rm cm}^{-3}\over n}~~{\rm years}.
\end{equation}
From this one can infer that accreted matter could effectively diffuse to other part from the polar cap.

When high-accretion process ceases, polar corona would not remain the same due to fusion and diffusion, but be converted from $\rho$-high to $\rho$-low phases.
The residual corona is cool, nonetheless, it could be heated by falling ions during a low accretion-rate phase (even via the Bondi accretion from interstellar medium), increasing the temperature $T_{\rm i}$ from $\sim$ keV to $\sim$ 10 MeV.
A higher ion temperature ($T_{\rm i}>T_{\rm e}$) of 4U 1700+24 in case of wind accretion could also be responsible for an emission (rather than an absorption) Ly-$\alpha$ line.

\subsection{The case of 4U 1700+24}

One of the symbiotic X-ray binaries, 4U 1700+24, is suggested to be a wind accretion system (Nucita et al. 2014), with X-ray luminosity $L_{\rm x}\sim 10^{34}$ erg/s and variability on both long- (months to years) and short-term timescales (10 to 1000 s).
We propose that 4U 1700+24 could be a strange quark-cluster star covered by a polar corona with a height of $h$.
Because the density of accreted matter is much lower than of the corona (see Eq.(\ref{rho-Bondi}) for an estimation), the detected redshift $z=0.009\pm 0.001$ of O VIII Ly-$\alpha$ emission line would hint a gravitational redshift, rather than the Doppler contributions due to the falling plasma velocity of Eq.(\ref{vf}).

\begin{figure}
\centering
  \includegraphics[width=0.6\textwidth]{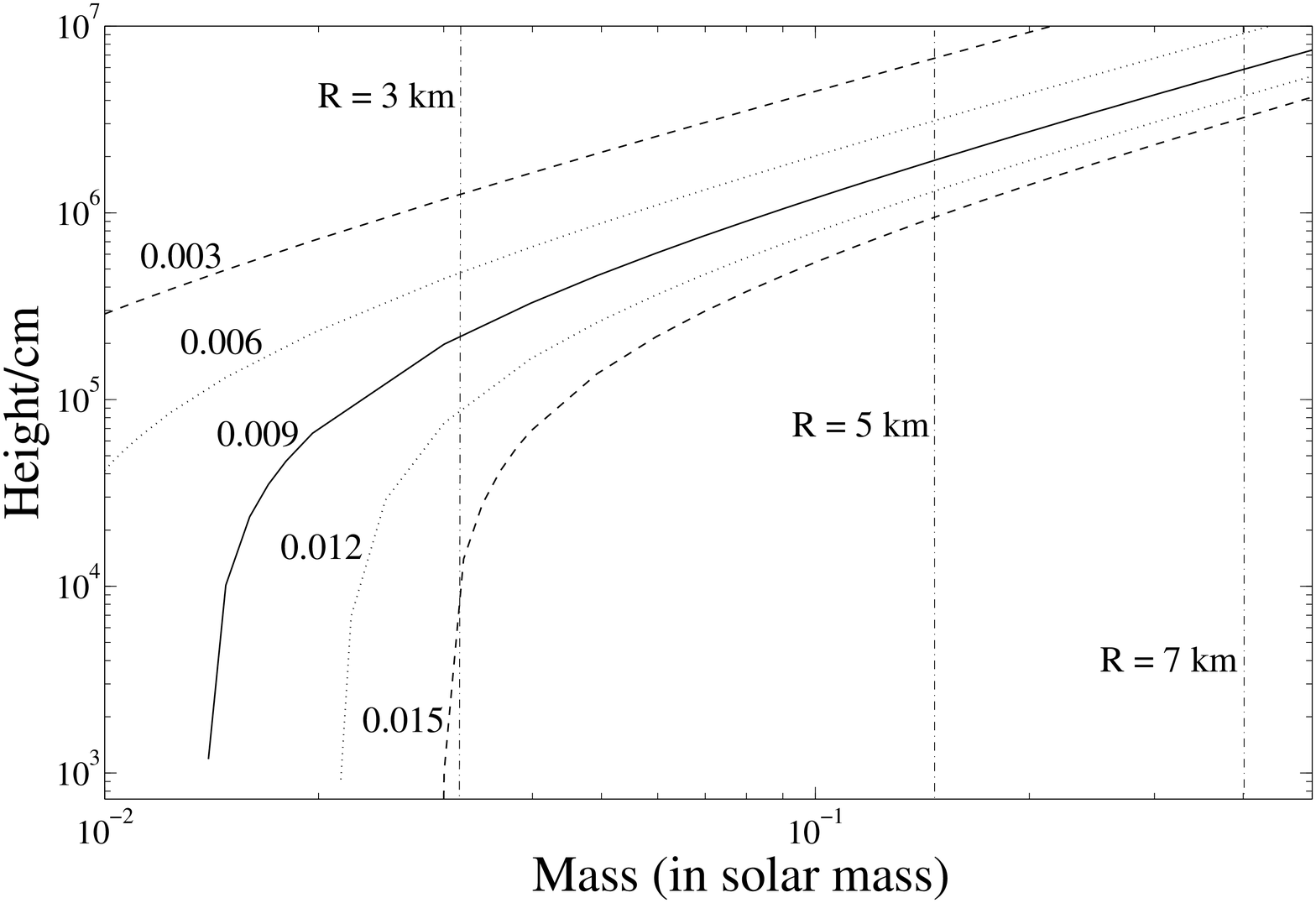}\\
  \caption{Contour lines of the redshift in mass-height diagram. Different redshifts ($z=0.003, 0.006, 0.009, 0.012, 0.015$) are labeled for five contour lines. The dash-dotted vertical lines ($R=3$ km, $R=5$ km, $R=7$ km) denote the radii of compact star as the radius could be easily calculated for a compact star with certain mass by Eq.(\ref{MR}).}%
  \label{fig.h-m}
\end{figure}

If $h\ll R$, one may infer a compact star with a few kilometers in radius (corresponding to $\sim 10^{-2}M_\odot$) from the redshift.
However, if the corona height is not negligible, the gravitational redshift should be modified to be
\begin{equation}
z\simeq {GM\over c^2(R+h)}.%
\label{z-h}
\end{equation}
Fig. \ref{fig.h-m} shows the implied corona height and stellar mass by the observed redshift, according to Eq.(\ref{z-h}). It is evident that 4U 1700+24 should be a low-mass quark-cluster star as long as $h<R$.

It is observed that the radius of the X-ray emitting region by the best-fit black-body model decreases as the overall X-ray luminosity decreases, from a few hundreds of meters down to a few tens of meters (Nucita et al. 2014).
This can also be well understood in the scenario presented.
From Eq.(\ref{rp}), one has the polar radius $r_{\rm p}\simeq 16$ m if $R=2.3$ km and $P=1$ s.
In a high accretion rate case, ion diffusion is significant so that polar hot spot is large, while in a low accretion rate case, the hot spot radius could approach the polar radius, $r_{\rm p}$.
We note that the hot spot radius could be much smaller than the corona radius although the corona could be as large as the whole star due to longtime diffusion.

\section{Conclusion and Discussion}

We propose that the compact object in the symbiotic X-ray binary, 4U 1700+24, is a low-mass quark-cluster star in a wind-accretion phase.
Because of very weak interaction to convert up or down to strange quarks, most of accreted ions are bounced back after collisions with quark-clusters, and accretion material through open-field lines may finally result in a formation of secular corona above the stellar surface, with stronger ion-diffusion at higher accretion rates.
Both the redshifted O VIII Ly-$\alpha$ line and the change of the black-body emitting region could be understood naturally in this scenario.


The nature of three kinds of isolated pulsars are focused in the literatures.
Their X-ray luminosity, $L_{\rm x}$, scales from $10^{30\sim 31}$ erg/s for the ``Magnificent Seven'', to $10^{32\sim 33}$ erg/s for central compact objects (CCOs), and to $10^{34\sim 35}$ erg/s for anomalous X-ray pulsars and soft $\gamma$-ray repeaters (AXP/SGRs).
The spin periods of CCOs are small, only a few hundreds of milliseconds, while that of other two are larger than a second, clustered around 10 s.
CCOs are proposed to be anti-magnetar (e.g., Mereghetti 2011) in order to explain their X-ray pulsations.
However, in the regime of quark-cluster stars, accretion from either inter-stellar medium or fossil disk could contribute significant heating on polar caps. Similar polar corona would also form there, with clearer pulsation for higher $L_{\rm x}$ (note: some of CCOs may be heated also by spindown powers).
In addition, the period clustering of AXP/SGRs and Magnificent Seven could also be understood with the inclusion of accretion torques (e.g., Liu et al. 2014).

It is conventionally believed that Type-I X-ray bursts are the results of unstable thermonuclear burning on the surface of neutron stars with binary accretion
(e.g., Lewin et al. 1993).
However, this popular point of view is challenged recently by a superburst detected in EXO 1745-248.
Altamirano et al. (2012) concluded that their observations of that superburst are very challenging for current superburst ignition models even assuming a few days of low-level accretion before the superburst onset.
Nevertheless, in the regime of solid quark-cluster star (e.g., Xu 2010), internal gravitational and elastic energy release during a quake of massive compact star might be responsible for triggering the bursts, manifesting as Type-I X-ray bursts.
Magnetic waves (e.g., Alfv\'en waves, Kaghashvili 1999) could effectively transport kinematic oscillation energy from the stellar interior out into the corona, heating significantly ions rather than electrons since the stellar oscillation frequency is much lower than the cyclotron frequencies of electrons and ions.
Nuclear flash accompanying a starquake could also be possible.
A recent study (Qu et al. 2014) of the flux evolution of three AXPs/SGRs does show that the ratios of persistent emission to the time averaged short bursts are comparable to those in the case of Type I X-ray bursts.
In any case, a future extensive study on this scenario is necessary.

\vspace{2mm}

\noindent
{\bf Acknowledgments}
This work is supported by National Basic Research Program of China (2012CB821800), NSFC (11225314) and XTP XDA04060604. An anonymous referee is sincerely acknowledged, especially for the language improving. I thank useful discussions at our pulsar group.

\label{lastpage}

\end{document}